\documentclass[a4paper]{article}

\usepackage[english]{babel}
\usepackage[latin1]{inputenc}
\usepackage[T1]{fontenc}
\usepackage{float}

\usepackage[a4paper,top=3cm,bottom=2cm,left=3cm,right=3cm,marginparwidth=2cm]{geometry}

\usepackage{amsmath}
\usepackage{algpseudocode}
\usepackage{graphicx}
\usepackage[colorinlistoftodos]{todonotes}
\usepackage[colorlinks=true, allcolors=blue]{hyperref}

\usepackage[numbers]{natbib}
\usepackage{url}

\begin{document}
\title{Defect patterns and software metric correlations in a mature ubiquitous system}

\author{Tim Hopkins\footnote{School of Computer Science, University of Kent, Canterbury, Kent, CT2 7NF, UK. t.r.hopkins@kent.ac.uk}, Les Hatton\footnote{SEC, Kingston University, KT1 2EE, U.K., lesh@oakcomp.co.uk}}

\maketitle

\begin{abstract}
Software engineering is not an empirically based discipline.  Consequently, many of its practices are based on little more than a generally
agreed feeling that something may be true.  Part of the problem is
that it is both relatively young and unusually rich in new and often
competing methodologies.  As a result, there is little time to infer
important empirical patterns of behaviour before the technology moves
on. Very occasionally an opportunity arises to study the defect growth and
patterns in a well-specified software system which is also well-documented
and heavily-used over a very long period.  

Here we analyse the defect growth and structural patterns in just such
a system, a numerical library written in Fortran evolving over a period
of 30 years.  This is important to the wider community for two reasons.
First, the results cast significant doubt on widely-held long standing language-independent beliefs and second, some of these beliefs are perpetuated in modern technologies.  It therefore makes good sense to use empirical long-term data as it becomes available to re-calibrate those generalisations.  Finally, we analyse the phenomenon of defect clustering providing further empirical support for its existence.

\end{abstract}

\section{Overview}
The ability to predict future program failures from static properties
of programs (i.e., properties which can be directly measured from the
source code) has long been a goal of software engineering researchers.
Efforts to predict future failures based on
error-prone language features (for example, mechanisms which lead to
loss of significant bits) have generally proven fruitful, see \cite{PflHat97}.
Unfortunately, efforts to find satisfactory correlations between
program failure and structural properties, such as the number and type of
decisions, are undermined by the very disparate nature of software with
a multitude of
programming languages employing many different paradigms.

Despite this, some beliefs have become surprisingly well-entrenched across
numerous languages in a wide range of application areas.  For example,
in spite of significant evidence to the contrary (see \cite{FentonNeil99,Shepperd88})
cyclomatic complexity \cite{McCabe76}, a graph theoretic measure
essentially related to the number of decisions within a piece of
computer software,
has been widely used in many disparate developments as a predictor of
components which will be unreliable in some sense.  The danger of its
continuing use is that it has become a design criterion for limiting
components to a maximum cyclomatic number as is the case in the
Joint Strike Fighter C++ standard, a safety-critical
environment, (\cite[AV Rule 3]{JSF05}).

Even longer standing is the debate over the \textit{goto} statement.
The result of this debate initiated by Dijkstra~\cite{Dijk68}
and unconstrained by any relevant measurement until very recently \cite{Nagappan18}, was that the
\textit{goto} statement is to this day believed to be strongly correlated
with program failure in whatever context it arises, including its implicit
forms such as the \textit{continue} and \textit{break} statement which
appear in numerous modern languages.  This appears in such influential
standards as the first two editions of MISRA C (\cite[Rule 56]{MISRA98},
\cite[Rule 14.4]{MISRA04}), and in no less than three forms in the
most recent incarnation \cite{MISRA12},
where the authors decided to downgrade the rule to advisory but added 
rules banning
forward jumps or jumps into a sub-block.  It also appears in
the Joint Strike Fighter C++ standards,
(\cite[AV Rule 189]{JSF05}), and the European Space Agency Ada standard
\cite{ESA_Ada98}.  Other language independent code fragments which have
invited opprobrium include the so-called dangling \textit{else if},
being an \textit{if} $\ldots$ \textit{else if} $\ldots$
clause with no \textit{else}
statement (\cite{JSF05} and \cite{MISRA04}), and also restrictions on
the maximum depth of nesting of control structures, \cite{ESA_Ada98}.

It is clear then that these and similar beliefs are well-entrenched in modern
development.  Indeed, they are so well-entrenched that they still appear in prosecution presentations and analyses in high-profile court cases such as the Toyota Unintended Acceleration Bug, \cite{NASA2011,Koopman2014} including pejorative discussions of the eponymous ``spaghetti'' code \cite{SRS2013}.  Some authors rightly question these apparently unassailable beliefs, \cite{Cummings2016}, but what is lacking is supporting evidence.  Related attempts to explore this space include learning algorithms for defect prediction \cite{FuMenzies17} but it is not clear how such abstract methods might help us to reject such deeply ingrained beliefs.

We depart from earlier defect studies here by directly analysing a number of such beliefs against a wide collection of metrics which software engineering has thrown up in the last three decades and show that none of these beliefs appear individually to have any statistically significant basis in fact when studied over a long period in a well-documented and well-specified system.  This is a crucial point.  No previous study has worked with a package of such maturity with a continuous defect history for the 30 years or so of its lifetime at the time this work was carried out.  \textit{If these beliefs had any substance, such effects should have appeared}.  This of course is not a guarantee but it is certainly persuasive.

Note that we do not seek to rehabilitate the \textit{goto} statement or indeed any of the programming fragments which have attracted negative comment over the years. However, we do strongly emphasise the importance of empirical
evidence in supporting claims if we are to understand the essence of
software engineering in order to improve its rather erratic relationship with
systems failure.  That such beliefs still appear in a modern context
and are accepted without empirical challenge is sufficient motive for
the present work.  The nature of design and component interactions does
indeed evolve with time as programming languages implement more or less
of such paradigms as OO, but all programmers must make decisions and such
decisions are often tainted with beliefs formulated in the distant past
without much intervening influence from measurement.

Not only are some practices condemned without measurement but others are
similarly supported without measurement.  Although more recent efforts
have produced interesting results for structural metrics in OO systems \cite{Subra03},
other empirical studies \cite{Hat98} suggest that some of the suggested
benefits of such systems may be rather more illusory than was hoped.
However, all such efforts rely on the availability of
high quality failure records over a period of time coupled with access
to source code and excellent version control.  Such opportunities do
not arise very often and here we are able to analyse one such dataset
acquired over an unusually long period and present the results using
standard statistical tests of significance as a contribution to the
empirical base of this subject.  The large number of components measured
give much confidence in the results while the relevance of these results
to the wider community is unquestionable for the reasons outlined above.

In the interests of both pedagogy and repeatable science, the
sanitised raw data and complete means to reproduce all the results of this work will be available for download and analysis\footnote{http://www.leshatton.org/} following the recommendations of \cite{Ince2012}.

\subsection{The analysis of a numerical software library} 

The NAG (Numerical Algorithms Group) Library~\cite{nag} is a very
widely-used set of scientific procedures.  Over the last thirty years,
they have been continually enhanced to keep pace with research in
numerical analysis and have also spread from the original implementation
language of Fortran 66 and  77 into other languages such as C, Ada and
Fortran 95.  Here we analyse the Fortran 77 library over a number of
releases which provides an excellent opportunity to study defect growth
for several reasons

\begin{itemize}
\item The package has a complete and carefully maintained defect history
   which was embedded in program headers and for which perl scripts to mine
   the header defect information could easily be designed.

\item The package is sufficiently large; 266,123 executable lines of code (XLOC)
   as analysed in 3659 subroutine/functions,

\item As is often the case with software experiments, no usage or coverage
   data was available but the data shown here covers a period of three decades,
   a relatively long maturity time and the defect density is likely
   to be more asymptotically representative.

\item The package is of good quality for its generation (1978 onwards) and
   covers a difficult application area with an asymptotic defect density of 4.9/KXLOC
   (thousand executable lines of code).

\item The package is unusually well-specified for a software system
   as it implements procedures defined in mathematical notation.  It is
   therefore much less vulnerable to the problems which occur in many systems
   through imprecisely defined requirements, \cite{Ince2012}.
   
\item Unlike later standardisations of Fortran, Fortran 77 has an incomplete set of iteration constructs and requires
   the programmer to construct them with the \textit{goto} statement, allowing an unusually detailed analysis of the impact of what is colloquially known as ``spaghetti code'' on the evolution of defect.

\end{itemize}

\subsection{Extraction of static measurements}
The complete source code of the library was made available to us and we
designed and implemented parsing tools for the full Fortran 77 language,
and tested them against the FCVS (Fortran Compiler Validation Suite).
This turned out to be necessary in order to be able to extract all the
desired static code measurements.  Given that most if not all previous work studying defect in software systems uses the somewhat ambiguous LOC (line of code) as an independent variable, it also gave us the opportunity to measure this against the more precise measure of programming language tokens as used by compilers to define programming languages.  We will expand on this shortly.

The parsing engine was designed using
hand-crafted lexical and syntactical analysers written in ISO C to cater for
Fortran features such as

\begin{itemize}

\item The generally non-significant behaviour of the space character
   in Fortran 77 (the first lexical step is to discard all spaces outside
   strings or in the first 5 character positions of a Fortran line).

\item The arbitrary nature of the look-ahead in Fortran necessary to
   resolve grammatical structures such as the I/O implied DO loop.

\end{itemize}

The code measurements, often known as metrics, were chosen on the basis of
their common occurrence in the literature or anecdotally over the years.
As a result, 17 properties were extracted from each of the 3659 components.
The five letter codes after each item header will be used as abbreviations
for the corresponding parameter throughout the rest of the paper.
We will continue to refer to these as metrics following industry practice whilst acknowledging that they do not necessarily preserve conventional mathematical definitions of a metric.  Shown alphabetically, they are

\begin{enumerate}
\item \textit{Number of backward jumps: STBAK}.  This is included for
     anecdotal reasons.  It is thought to interfere with readability.

\item \textit{Cyclomatic complexity: STCYC}. This is a  graph theoretic
     measurement which is essentially a count of the number of 
     decisions within a software component, It was first introduced by 
     McCabe~\cite{McCabe76}.
    
\item \textit{Number of dangling elseifs, i.e., an if $\ldots$ else if with
     no else clause: STELF}.  This is included for anecdotal reasons and is
     believed to indicate the presence of incomplete logical thought.

\item \textit{Number of gotos: STGTO}.  There has been so much discussion
     of this over the years since the initial comments in~\cite{Dijk68},
     that we felt we could not leave it out.  In addition, Fortran 77
     has a rich set of goto forms including the arithmetic \textit{if}
     and the absence of any form of WHILE construct means that goto
     statements in various forms are used unusually frequently so the current study is particularly suitable to illuminate this.
     
\item \textit{Knot count: STKNT}.  A knot is a crossing of control flow as
     illustrated, for example, by Shooman~\cite{Shoo85}.  Knots only occur in
     languages which have explicit non-structural jump constructs such
     as the eponymous goto statement in its various forms.  The goto
     statement is a necessary but not a sufficient condition for a knot
     as it can be used to simulate nested (knot-free) structures as well
     as non-nested structures.  The existence of knots is often referred
     to informally as `spaghetti' code.
     
\item \textit{Source lines of code: STLIN}.  A simple count of the number
     of source lines of code including comment as would be seen in a 
     text editor.  See also STXLN.
     
\item \textit{Extended cyclomatic complexity: STMCC}.  This is 
     an extension to cyclomatic complexity based on complex 
     predicates~\cite{Myers77}.  A complex predicate contains either
     logical disjunctive (`or') or conjunctive (`and') phrases or both.
     
\item \textit{Maximum level of nesting of \textit{if} statements: STMIF}.
     This is included for anecdotal reasons.  It is thought to be associated
     with testing difficulties.
     
\item \textit{Total number of unique operands: STOPN}.
     See STOPT.
     
\item \textit{Total number of unique operator tokens: STOPT}.
     A recent look at the application of information theory suggests
     STOPN and STOPT together play a pivotal role, \cite{HatTSE14}.

\item \textit{Path count: STPTH}.  The path count is the
     number of ways through a particular program assuming that all paths
     are equally likely \cite{SaferC}.  The rationale behind it is that
     it is more sensitive to decision complexity than the cyclomatic
     number (for example, it can distinguish between a sequential series
     of if statements and a single switch statement containing the same
     number of clauses which have the same cyclomatic complexity).
     This is similar to the NPATH metric put forward in~\cite{Nej88}.

\item \textit{Number of subroutines in a file: STSUB}.  This is included
     for anecdotal reasons, however, it should be noted that in Fortran 77,
     unlike C, the file has no special linguistic meaning.
     
\item \textit{Total number of global variables referenced: STTCM}.
     In Fortran Global variables are shared through the blank and
     named common construct and such variables are shared across all
     functions referencing those common areas.
     
\item \textit{Total number of operator and operand tokens: STTOT}.
     See STOPT.

\item \textit{Number of undeclared variables: STUNV}.  There has
     again been largely anecdotal attribution that this indicates some
     level of sloppiness and may therefore be related to defect.
     
\item \textit{Number of declared objects actually used: STVAR}.  This is
     included for anecdotal reasons.
     
\item \textit{Number of executable lines of code: STXLN}.  Executable
     lines of code (XLOC) is a count of the number of lines which generate
     executable code when compiled.  Many defect models have been built
     using executable or one of the other measures of lines of code as
     an independent variable (see, \cite{Lip82}, \cite{BasiliPerricone1984},
     \cite{Koru2007} and many others). The main attraction of using lines
     of code is that they are usually very easy to measure although
     they have a built in sensitivity to layout.   Note that Fortran
     continuation lines are included but not separately counted.

\end{enumerate}

\section{Statistical analysis}

The classic approach in studies like this has been to look for a positive correlation between defect  and XLOC or the appearance of a particular programming construct such as the eponymous \textit{goto} statement, or the presence of \textit{if} $\ldots$ \textit{else if}
statements without an \textit{else} clause, (the so-called dangling
else construct), or the cyclomatic number.  The object of such studies is to develop rules,
particularly for safety-critical systems, such that the overall level of defect
might be reduced by avoiding such constructs.

\subsection{Methodology}
To support full reproducibility, the dataset, the experiments and the complete means of analysis are all provided with or referenced by this paper.  The extraction of the data has already been described in detail in earlier sections.  The data was analysed using R with the following parameters:-

\begin{verbatim}
R version 3.2.3 (2015-12-10)
Copyright (C) 2015 The R Foundation for Statistical Computing
ISBN 3-900051-07-0
Platform: i686-pc-linux-gnu (32-bit)
\end{verbatim} 

We will start then by investigating
pairwise correlations in the raw data.  The complete dataset
consists of 3,659 data points, one for each subprogram for which a complete set
of metrics was extracted as defined above.

\subsection{Pairwise linear correlations with defects}
For each metric, MMMM, the raw data was handled by the following R
program, in which the intercept is forced to be zero, which seems
reasonable.

\begin{verbatim}
df <- read.csv(file="defect_MMMM.csv",head=TRUE)
plot(df)
y <- df$defect
x <- df$MMMM
fm=lm(y~x+0,data=df)
summary(fm)
\end{verbatim} 

The results for the pairwise linear model of defect against each metric
MMMM are given in Table~\ref{table:rawpw}.

\begin{table}[ht]
\centering
\begin{tabular}{c c c c}
Metric & F-stat & Adj. $R^{2}$ & p \\
\hline
STBAK & 423.2 & 0.104 & $< 2.2.e^{-16}$ \\
STCYC & 888.5 & 0.195 & $< 2.2.e^{-16}$ \\
STELF & 101.6 & 0.027 & $< 2.2.e^{-16}$ \\
STGTO & 848.6 & 0.188 & $< 2.0.e^{-16}$ \\
STKNT & 348.2 & 0.087 & $< 2.0.e^{-16}$ \\
STLIN & 978.3 & 0.211 & $< 2.0.e^{-16}$ \\
STMCC & 505.1 & 0.121 & $< 2.0.e^{-16}$ \\
STMIF & 316.2 & 0.080 & $< 2.0.e^{-16}$ \\
STOPN & 1017 & 0.218 & $< 2.0.e^{-16}$ \\
STOPT & 796.1 & 0.179 & $< 2.0.e^{-16}$ \\
STPTH & 578.5 & 0.136 & $< 2.0.e^{-16}$ \\
STSUB & 555.9 & 0.132 & $< 2.0.e^{-16}$ \\
STTCM & 522.1 & 0.125 & $< 2.0.e^{-16}$ \\
STTOT & 946.4 & 0.206 & $< 2.0.e^{-16}$ \\
STUNV & 390.7 & 0.097 & $< 2.0.e^{-16}$ \\
STVAR & 1239 & 0.253 & $< 2.0.e^{-16}$ \\
STXLN & 913.9 & 0.200 & $< 2.0.e^{-16}$ \\
\hline
\end{tabular}
\caption{Raw data pairwise linear correlations for metrics against defects}
\label{table:rawpw}
\end{table}

The confidence intervals for the fitting coefficients are not shown for
the simple reason that although the fits are highly significant as can be
seen by the high F-value and low p, the adjusted R-squared value reveals
that individually very little of the variance is explained by any single metric.  \textit{In other
words, they are effectively useless as individual predictors of defect.}

\begin{figure}[h]
\centering
\includegraphics[width=10cm,height=8cm]{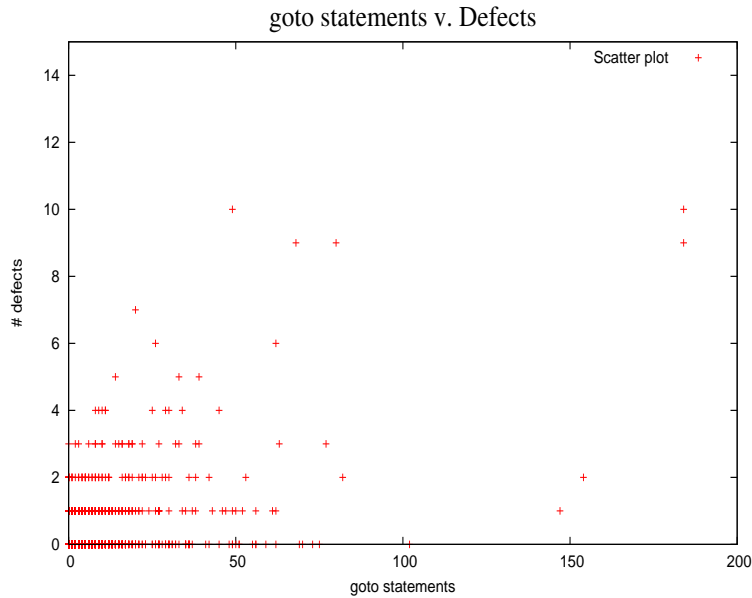}
\caption{A scatter diagram between the number of defects and the number
of goto statements.}
\label{fig:goto}
\end{figure}

To convey the very noisy nature of these data, the pairwise
correlation of defects against number of goto statements is shown
in Figure~\ref{fig:goto}.  This graph is typical of the unsmoothed
relationship between defects and most of the metrics in this study.

Given the modest contribution to variance of any individual metric,
Principle Component Analysis was then performed on the raw data to
explore any possible relationship with linear combinations of metrics
revealed by rotating the data cloud.

\subsection{Principle Component Analysis}
The first step was to compute the variances of the principle components,
scaling the data because of the widely different scales of each metric
and produce a biplot using the following R script.

\begin{verbatim}
## Read in the raw data.
pcadf <- read.csv(file="pca_defect.csv")
pca <- prcomp(pcadf, scale. = TRUE)
## Plot the eigenvalues
setEPS()
postscript("pca_evalues.eps")
plot(pca)
dev.off()

postscript("pca_biplot.eps")
biplot(pca)
dev.off()
\end{verbatim} 

\begin{figure}[h]
\centering
\includegraphics[width=10cm,height=8cm]{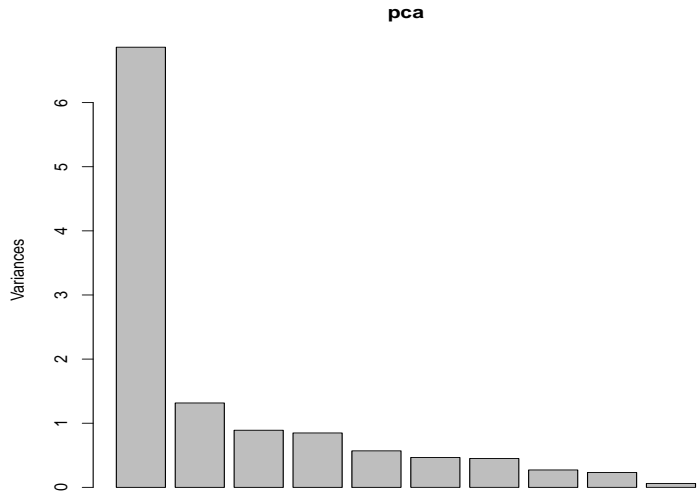}
\caption{The variances of the Principle Components.}
\label{fig:pca_evalues}
\end{figure}

The result is shown as Figure~\ref{fig:pca_evalues}.  As is usually
the case with PCA, most of the variance is explained by a small number
of components.  The `knee' after the first component is not particularly well defined, and we considered only the first two noting that the first four account for around 83\% of the total variance.

\begin{figure}[h]
\centering
\includegraphics[width=14cm]{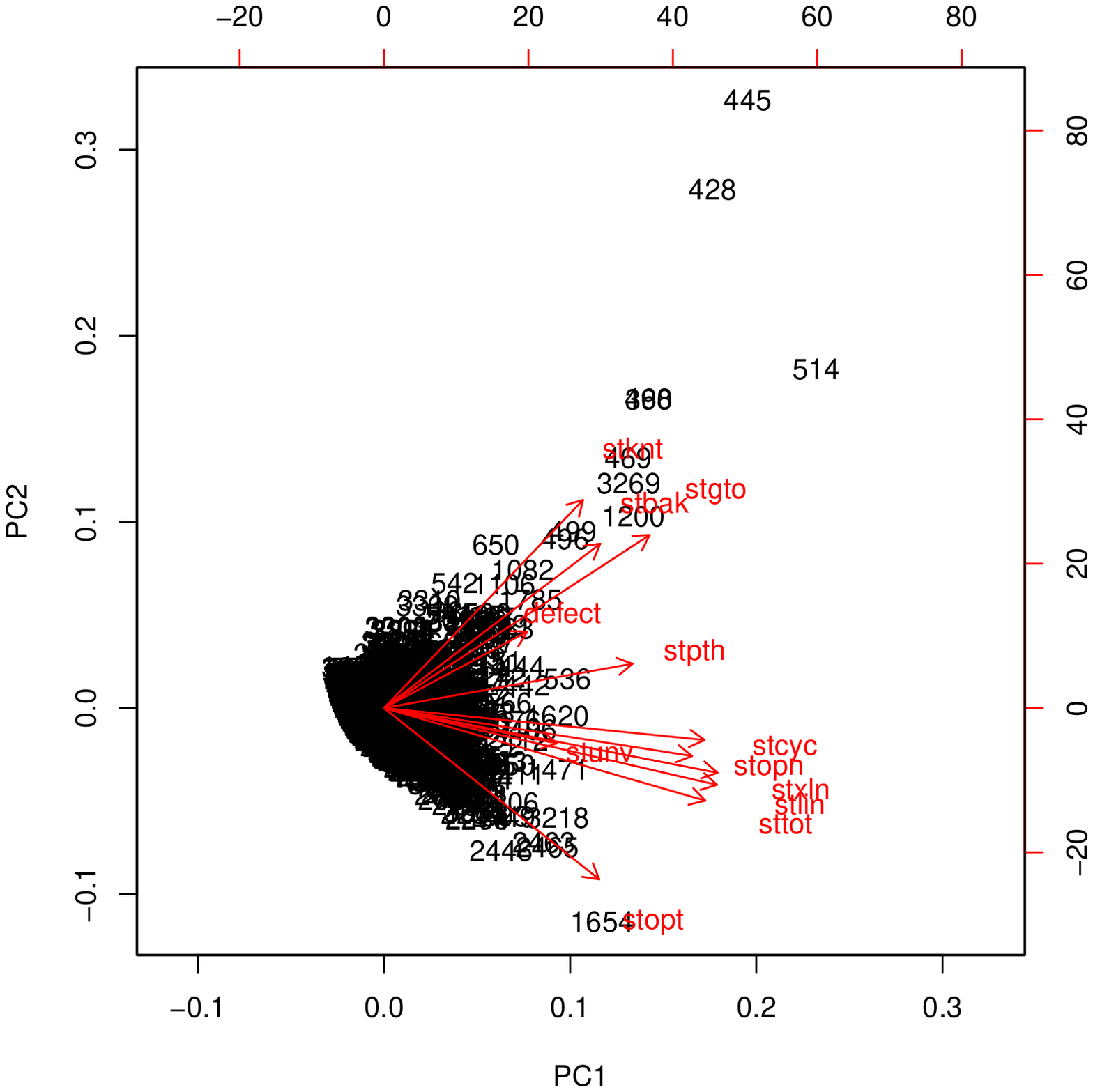}
\caption{A biplot of the contributions to the two principle components.}
\label{fig:pca_biplot}
\end{figure}

The resulting biplot is shown as Figure~\ref{fig:pca_biplot}.  The
numbers in black are the rows from the data matrix, whereas the vectors
show the contributions and directions with respect to the first two
principle components.  Whilst generally pointing in the same direction,
the biplot shows several clusters.  The most significant cluster is
essentially related to size and groups STCYC, STOPN, STXLN, STLIN,
STTOT and to a lesser extent STUNV\@.  A secondary cluster more related to
perhaps departures from nested structure, groups STKNT, STBAK and
STGTO\@.
The relationship with defect is however clearly complicated.

We conclude that in the NAG dataset, \textit{there is no simple linear
relationship in the raw data between single parameters and defect
growth which unequivocally supports the nature of the rules given in
the coding standards described earlier even though the dataset is of
sufficient maturity that by now, any such phenomenon, if present, would
be expected to have manifested itself significantly}.  Instead, the
predominant variability in the data cloud is aligned along a direction
which is defined by a complex and relatively evenly balanced combination
of several parameters which would appear to defy any attempt at
rendering into a simple and intuitively justifiable replacement rule.

\begin{figure}[h]
\centering
\includegraphics[width=14cm]{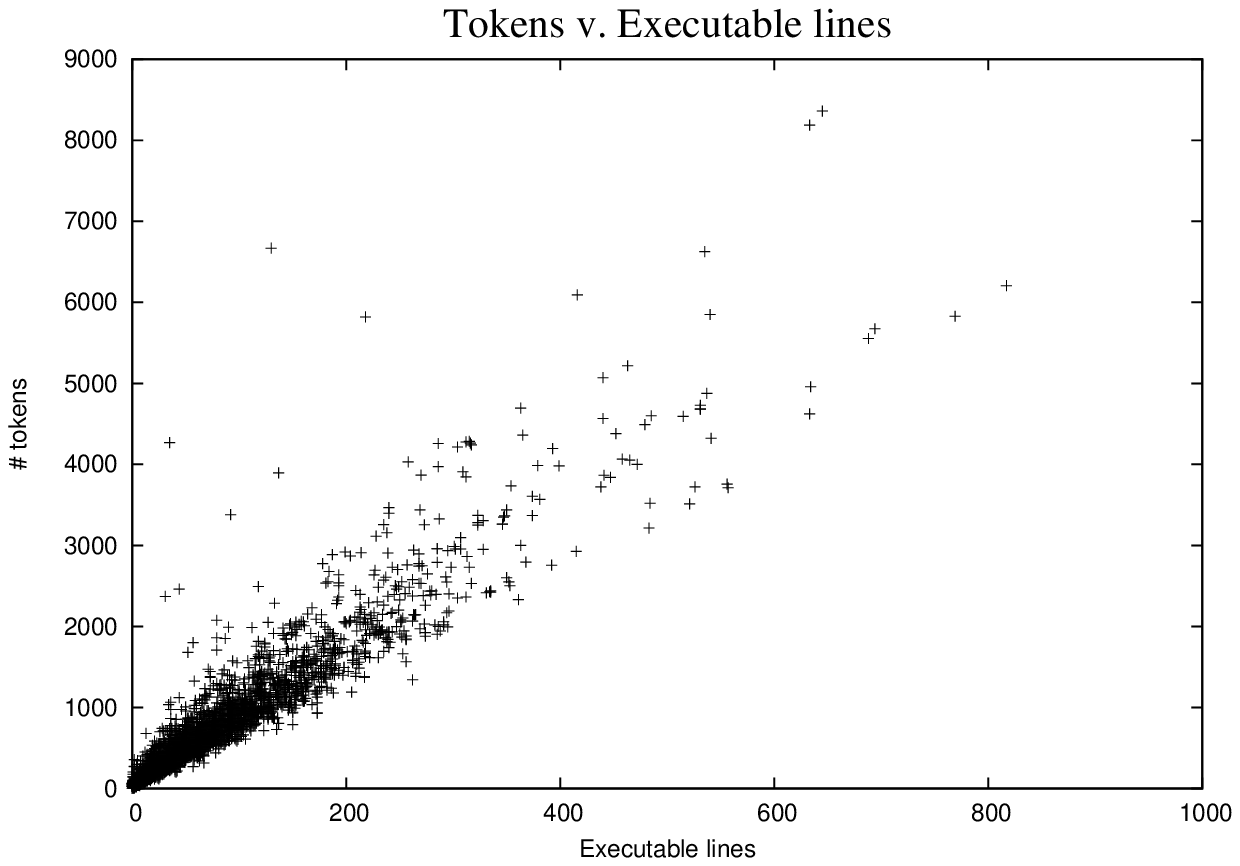}
\caption{A correlation plot of the number of executable lines (XLOC) against the actual number of programming tokens.}
\label{fig:lntok}
\end{figure}

The general similarity of data for each metric suggests the possibility
that pairwise combinations of them may be individually highly correlated
and we will now investigate this.  Since XLOC have played such a part
in the historical development of defect analysis in software systems and we are investigating empirical rules which have evolved around using lines of code, we will confine ourselves to studying correlations against XLOC, although we note that the behaviour of a program is defined by its \textit{tokens} not its lines of code, (which effectively have no meaning in a language compiler).  To emphasize this, Figure \ref{fig:lntok} shows the correlation between XLOC and tokens for each function and subroutine in the NAG library, indicating that they are as expected, highly correlated, but they are not the same and recent studies in the length distribution of software components confirms that the difference can be important, \cite{HatTSE14}. 

\subsection{Cyclomatic complexity versus XLOC}
\begin{figure}[h]
\centering
\includegraphics[width=10cm,height=8cm]{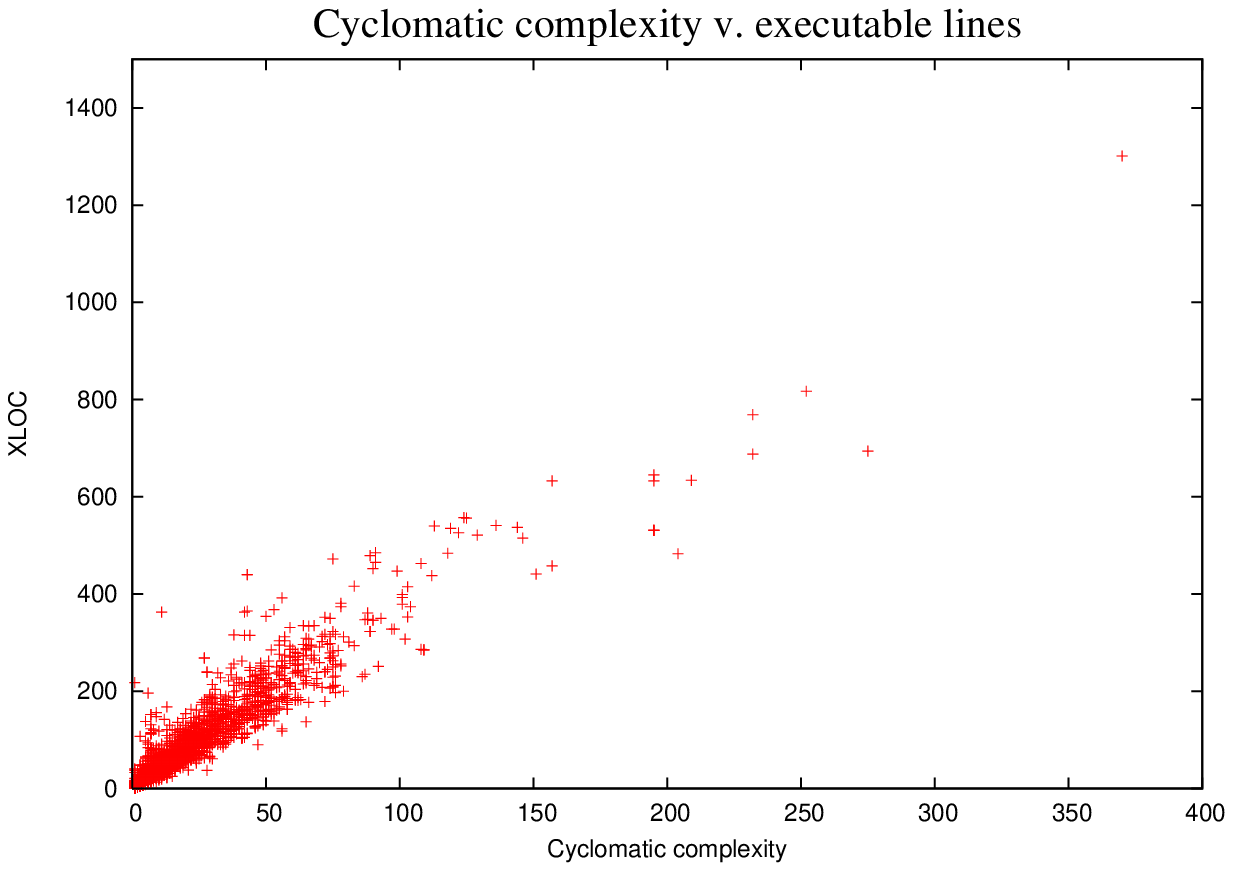}
\caption{A scatter diagram between the number of executable lines and
the cyclomatic complexity.  A strong correlation is clearly visible in which nearly all the variance is explained by a linear relationship
between the two.}
\label{fig:cycxl}
\end{figure}

A short analysis revealed that cyclomatic complexity is very
highly correlated with executable lines of code as shown in
Figure~\ref{fig:cycxl}.

Taking the reasonable measure of forcing a zero
intercept, yields a regression equation of

\begin{equation}
XLOC = 3.96 \times STCYC 
\end{equation} 
The F-statistic in this
case is 45200 and the adjusted R-squared value is 0.9251 with $p <
2.0.e^{-16}$. In essence, then this equation states that it is extremely
likely that there will be a decision about every 4 executable lines in a
typical program in this library, a perhaps not unsurprising observation.
This has been noted numerous times, e.g. \cite{BasiliPerricone1984}
and also very recently and emphatically across different languages by
\cite{Meulen08} on a very much larger sample.  Here we simply confirm
the observation that \textit{the cyclomatic complexity appears to add
little if any significant information to that already contained in the
count of executable lines} and the two are almost interchangeable, rather
like comparing somewhat noisy temperature measurements in both Fahrenheit
and Centigrade.

There is nothing therefore in this study which offers support for
standard software engineering folklore relating conventional software metrics with defect in a system of sufficient maturity that any such relationships should by now have appeared at some level of significance.  It is reasonable to
ask then if there are \textit{any} useful patterns related to defect which
can be identified.

\section{Defect clustering in the NAG Fortran library}
In the NAG Fortran library, defects were distributed by size and
by executable lines of code as shown in Table~\ref{tab:Fdefects}.  This is most interesting.
The tendency for defects to cluster is particularly prevalent here, with all of the defects so far seen occurring in only 20\% of the components, and very close to 80\% of them occurring in only 20\% of the lines.  These data are therefore entirely consistent with
the observations of \cite{BoehmBas2001}.

\begin{table}[h]
\centering
\begin{tabular}{|c|c|c|}
\hline Number of defects & Number of components & XLOC \\
\hline
\hline 0 & 1749 (80.19\%) & 120632 (71.97\%) \\ 
\hline 1 & 322 (14.76\%) & 31215 (18.62\%) \\ 
\hline 2 & 65 (2.98\%) & 8033 (4.79\%) \\ 
\hline 3 & 24 (1.10\%) & 3173 (1.89\%) \\ 
\hline 4 & 10 (0.46\%) & 1401 (0.84\%) \\ 
\hline 5 & 3 (0.14\%) & 507 (0.30\%) \\ 
\hline 6 & 2 (0.09\%) & 684 (0.41\%) \\ 
\hline 7 & 1 (0.05\%) & 111 (0.07\%) \\ 
\hline 8 & 0 (0.00\%) & 0 (0.00\%) \\ 
\hline 9 & 3 (0.14\%) & 1122 (0.67\%) \\ 
\hline 10 & 2 (0.09\%) & 746 (0.45\%) \\ 
\hline 
\end{tabular}
\caption{Defect counts v. subroutines and executable lines of
code in the NAG Fortran library.}
\label{tab:Fdefects}
\end{table} 

At first sight, this table appears to open up forensic possibilities whereby there might be specific reasons why 80\% of the components have exhibited no defect in three decades of heavy use.  However, the previous studies indicate that if there are, none of the metrics or metric combinations can explain this.  As a simple example, perusing Figure \ref{fig:goto} reveals that components containing anywhere between 0 and around 190 gotos exhibited zero defect with no obvious pattern.

\subsection{Conditional probabilities of defect detection}
The tendency for defects to cluster may however shed light on optimal defect searching strategies \cite{Koru2007}. Specifically, if a component is known to contain a defect, does it make sense to look for more defects in the same component or to look elsewhere?
To resolve this we can cast the data of Table \ref{tab:Fdefects} into a simple
conditional probability.

\begin{figure}[htb]
\centering
\includegraphics[width=10cm,height=8cm]{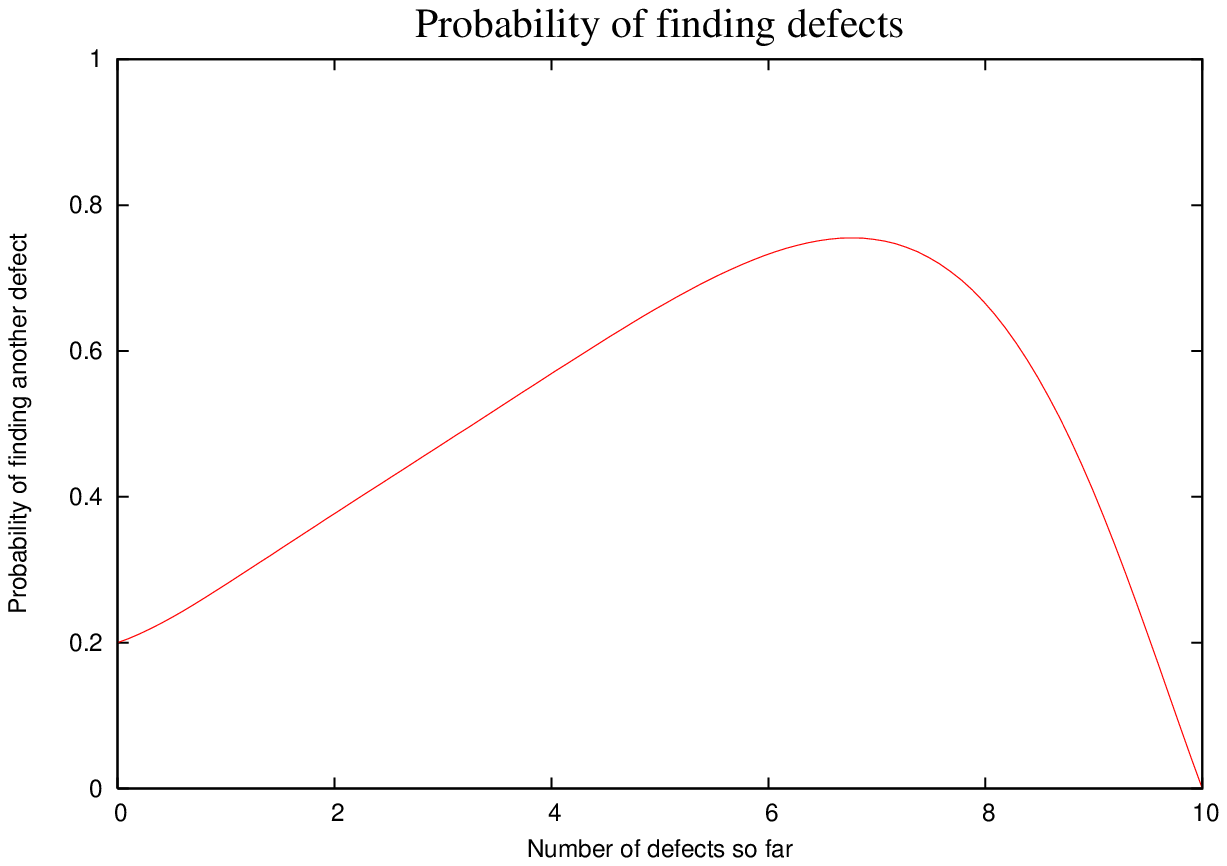}
\caption{The probability of finding an $(N+1)$th defect in a subroutine/function in the NAG
Fortran library, given that $N$ defects have been found so far.}
\label{fig:fcondprob}
\end{figure}

Let $P(N+1|N)$ be the probability of finding the $(N+1)$th defect in
a component given that $N$ have been found already in that component.
As an example, for Table \ref{tab:Fdefects}, the conditional probability
of finding 4 defects given that we have found 3 already is given by

\begin{equation}
P(4|3) = \frac{10+3+2+1+0+3+2}{24+10+3+2+1+0+3+2} = 0.47
\end{equation} 

which is a surprisingly high value.  Figure \ref{fig:fcondprob} shows the
respective probabilities $P(N+1|N)$  as a function of $N$. 
It is very clear that up to quite a high number of defects, the conditional probability of finding further defects in the same subroutine/function counter-intuitively continues to increase.

\section{Discussion and Conclusions}
A large and widely used mature scientific subroutine library has been
analysed to test a number of widely-held beliefs about the relationship
between defects and either the structural properties of the code or its
language features.  The relevance of this to the wider community is that
such conjectures have been held for many years and are widely applied in
modern programming standards even in critical systems, and may therefore
influence the failure behaviour of such systems.  These include
the relationship of defects with the \textit{goto} statement, dangling
\textit{else} constructs, number of decisions, maximum levels of decision nesting, component
size, number of globally accessible variables, and a number of other code features which occur frequently in programming standards \textit{for numerous programming languages.}

We conclude the following

\begin{itemize}
\item No single metric of those analysed here has any statistically significant relationship with the occurrence of defect in the thirty years of use of this ubiquitous software package.  In view of their persistence in the folklore and consequent appearance in court proceedings relating to software failure, we single out the \textit{goto} statement, the cyclomatic complexity, the number of globally accessible variables and measurements of spaghetti code such as \textit{knots} as being \textit{statistically unrelated to the appearance of defect in this package.}

\item Principle Component Analysis shows that although most of the variance can be explained by a single linear combination of individual metrics, this does not correspond to any simple intuitively useful rule.  Rather, the relationship is complex and suggestive of mere data-fitting.  This would be particularly true with this dataset if non-linear relationships between metrics were considered, defeating the central object of this paper which is to challenge a number of well-entrenched and incorrect views.
\end{itemize}

In contrast to these negative conclusions,

\begin{itemize}
\item Very strong evidence of defect clustering was found with typically 80\% of all components in this package exhibiting no defect in the entire measured life-cycle.  The metrics themselves throw no light on why some components develop defect whilst the majority do not, from which we suspect that the effect is essentially a manifestation of randomness in some form.
\end{itemize}

We stated with evidence, at the beginning of this paper, that a number of long-held beliefs have found their way into modern standards for development across multiple languages, even for safety-critical systems.  These have
inevitably influenced the way that developers have produced systems so it
is of some considerable importance to underpin them with empirical
support wherever possible.  In the present study on a mature system
over many years, a number of those beliefs, if supported, \textit{should have
left identifiable traces at some level of significance.}  They have not.

This is just one study but it is large, covers a long period and was
carried out in an application area which was unusually well-specified
implying that the resulting defect data is of higher precision than in
less well-specified areas.  Furthermore, the beliefs which were investigated are essentially programming language independent so we would have expected some manifestation of them if they were indeed valid.  Perhaps the biggest single lesson therefore is that beliefs which seem reasonable but are unsupported by any empirical evidence simply can not be trusted. 

By casting doubt on these beliefs, we do not seek to rehabilitate bad practice, but if we are to make progress in avoiding failures with modern technologies, we need to be able to quantify what ``bad practice'' actually means, and only those methodologies soundly based on empiricism are likely to be of any lasting help.

This is particularly apposite today where folklore appears to have become an intrinsic part of the forensic process of judging a software system's quality, potential for failure and indeed potential legal liability.

Finally, this work was completed in around 2005 when the NAG library was approximately 30 years old.  It has hung around as a departmental report since at the University of Kent Computing Laboratory but given that software engineering still has a dearth of reliable empirical data, it has been cleaned up here and cast into the context of more recent work.  Its message is however undimmed by the passage of time.

\bibliographystyle{unsrtnat}

\end{document}